\documentclass[sigconf,nonacm]{acmart}
\usepackage{makecell}

\AtBeginDocument{%
  }

\copyrightyear{2026}
\acmYear{2026}
\setcopyright{cc}
\setcctype{by}
\acmConference[CHI EA '26]{Extended Abstracts of the 2026 CHI Conference on Human Factors in Computing Systems}{April 13--17, 2026}{Barcelona, Spain}
\acmBooktitle{Extended Abstracts of the 2026 CHI Conference on Human Factors in Computing Systems (CHI EA '26), April 13--17, 2026, Barcelona, Spain}
\acmDOI{10.1145/3772363.3798440}
\acmISBN{979-8-4007-2281-3/2026/04}

\begin{document}

\title{When Friction Helps: Transaction Confirmation Improves Decision Quality in Blockchain Interactions}

\author{Eason Chen}
% \email{EasonC13@cmu.edu}
\affiliation{
  \institution{Carnegie Mellon University}
  \city{Pittsburgh}
  \state{Pennsylvania}
  \country{USA}
}

\author{Xinyi Tang}
\affiliation{
  \institution{Carnegie Mellon University}
  \city{Pittsburgh}
  \state{Pennsylvania}
  \country{USA}
}

\author{George Digkas}
\affiliation{
  \institution{Mysten Labs}
  \city{Palo Alto}
  \state{California}
  \country{USA}
}

\author{Dionysios Lougaris}
\affiliation{
  \institution{Mysten Labs}
  \city{Palo Alto}
  \state{California}
  \country{USA}
}

\author{John E. Naulty Jr}
% \email{jnaulty@mystenlabs.com}
\affiliation{
  \institution{Mysten Labs}
  \city{Palo Alto}
  \state{California}
  \country{USA}
}

\author{Kostas Chalkias}
% \email{kostas@mystenlabs.com}
\affiliation{
  \institution{Mysten Labs}
  \city{Palo Alto}
  \state{California}
  \country{USA}
}

\begin{abstract}
In blockchain applications, transaction confirmation is often treated as usability friction to be minimized or removed. However, confirmation also marks the boundary between deliberation and irreversible commitment, suggesting it may play a functional role in human decision-making. To investigate this tension, we conducted an experiment using a blockchain-based Connect Four game with two interaction modes differing only in authorization flow: manual wallet confirmation (\textbf{Confirmation Mode}) versus auto-authorized delegation (\textbf{Frictionless Mode}).
Although participants preferred \textbf{Frictionless Mode} and perceived better performance ($N=109$), objective performance was worse without confirmation in a counterbalanced deployment (Wave~2: win rate $-11.8$\%, $p=0.044$; move quality $-0.051$, $p=0.022$). Analysis of canceled submissions suggests confirmation can enable pre-submission self-correction ($N=66$, $p=0.005$).
These findings suggest that transaction confirmation can function as a cognitively meaningful checkpoint rather than mere usability friction, highlighting a trade-off between interaction smoothness and decision quality in irreversible blockchain interactions.

\end{abstract}

\begin{CCSXML}
<ccs2012>
   <concept>
       <concept_id>10002978.10003029.10011703</concept_id>
       <concept_desc>Security and privacy~Usability in security and privacy</concept_desc>
       <concept_significance>500</concept_significance>
       </concept>
   %     <concept_id>10003120.10003121.10011748</concept_id>
   %     <concept_desc>Human-centered computing~Empirical studies in HCI</concept_desc>
   %     <concept_significance>500</concept_significance>
   %     </concept>
 </ccs2012>
\end{CCSXML}

\ccsdesc[500]{Security and privacy~Usability in security and privacy}
% \ccsdesc[500]{Human-centered computing~Empirical studies in HCI}

\keywords{blockchain usability, authorization automation, usable security, decision quality}



\maketitle

\section{Introduction}

Blockchain applications provide guarantees such as integrity, non-repudiation, and tamper resistance \cite{blackshear2024sui}. In practice, these guarantees are enacted through user-facing authorization steps. Even for routine actions, users must interpret wallet prompts, verify transaction details, and explicitly confirm submission. Classic security-usability work shows that when security interfaces conflict with users' mental models, users often make errors or fail to complete intended security tasks, even when the interface appears usable by conventional standards \cite{whitten1999johnny,chen2025suigpt}. In cryptocurrency ecosystems, wallet usability problems are widely reported and can lead to frustration, disengagement, and irreversible mistakes \cite{voskobojnikov2021u,albayati2021study}. These persistent user experience barriers motivate a major design direction in web3, namely, reducing friction by automating authentication and authorization.

Recent innovations have made transaction authorization increasingly less visible to users. Modern wallet infrastructures and account abstraction mechanisms, such as zero-knowledge login schemes \cite{baldimtsi2024zklogin,chen2024buck} and embedded wallet frameworks like Privy \cite{privy2026}, can provide a temporary, in-browser signing capability, such as an ephemeral private key or session key stored locally, with one-time or time-bounded authorization.  Once such a capability is established, users can interact with applications without repeatedly switching to an external wallet interface for signature confirmation, enabling interaction flows that more closely resemble conventional web applications. For example, \href{https://pump.fun}{Pump.fun}, one of the top memecoin trading platforms, encourages users to use Privy login for ``zero confirmation trading''.

Despite their growing adoption, the behavioral consequences of removing explicit authorization steps remain poorly understood. Prior research on automation suggests that reducing manual intervention can improve efficiency and subjective experience, while also altering monitoring behavior and decision strategies \cite{parasuraman1997humans,parasuraman2010complacency}. Trust in automated systems can further amplify reliance, even when such reliance is unwarranted \cite{dzindolet2003role}. At the same time, interruption research has shown that context switches can impair goal maintenance and increase error rates in structured cognitive tasks \cite{altmann2014momentary,monk2004recovering,hodgetts2006interruption}, suggesting that eliminating wallet confirmations should improve both user experience and task performance. However, explicit confirmation dialogs may also be interpreted as a final opportunity for verification and reconsideration, particularly in irreversible decision contexts such as trading and gaming, as well as in scenarios where blockchain transactions may encode contract-like or legally meaningful obligations \cite{roche2021ergo,chen2023conversion}.

We therefore ask: \emph{How does removing explicit transaction confirmation affect human decision-making quality in blockchain interactions?} We examine authorization flow as a ubiquitous checkpoint in blockchain interactions and test whether eliminating it changes objective decision quality in a repeated, irreversible decision setting. As a controlled first step, we isolate this question using a game-based task that maximizes internal validity by holding all factors constant except the authorization flow, deferring ecological validation to higher-stakes domains as future work (see Section~\ref{subsec:ecological}). We implement a blockchain-based Connect Four game on the Sui network with two interaction modes that differ only in authorization flow. In the \textbf{Confirmation Mode}, each move requires a manual wallet confirmation. In the \textbf{Frictionless Mode}, moves are automatically authorized on the user's behalf, eliminating the confirmation step and creating a smoother interaction flow.
To sum up, our study yields two key contributions:

\begin{itemize}
  \item \textbf{A dissociation between perceived and objective performance under authorization automation.}  
We show that removing transaction confirmation increases subjective usability and perceived performance, yet systematically decreases objective performance, revealing a misalignment between users' self-assessment and actual decision quality in frictionless blockchain interactions.

\item \textbf{Transaction confirmation as a potential cognitive checkpoint.}  
Behavioral traces suggest that confirmation dialogs may serve as a reflective moments, such as pre-submission reconsideration and self-correction, as evidenced by canceled and revised moves.

\end{itemize}

\section{System Design and Development}
\label{sec:system}

To study how transaction authorization mechanisms influence human decision-making, we implemented a blockchain-based Connect Four game on the Sui network \cite{blackshear2024sui,chen2023building}. The system consists of a web-based frontend, an on-chain smart contract that enforces game rules, and an AI engine that both acts as the opponent and supports the evaluation of human decisions. A backend service orchestrates gameplay and logging.

Due to page limitations, complete implementation details like source code and screenshots of the UI are provided in an anonymous GitHub at:
\url{https://anonymous.4open.science/r/game-connect-four-study-E328/README.md}.

\subsection{Hybrid AI Engine for Gameplay}
The AI engine combines Monte Carlo Tree Search (MCTS) \cite{kocsis2006bandit} to generate adaptive opponent behavior and Minimax search with Alpha Beta pruning \cite{knuth1975analysis} for deterministic state evaluation. This hybrid design decouples stochastic gameplay from deterministic evaluation, allowing realistic opponent interactions while keeping evaluation independent of opponent randomness.
MCTS governs the AI opponent's moves and produces adaptive, non-deterministic play that approximates human-like behavior. In parallel, Minimax is used to compute deterministic evaluations of board states and enumerate all legal actions available to the human player. This separation ensures that opponent behavior remains engaging without entangling evaluation with stochastic gameplay dynamics.

\subsection{Adaptive Difficulty Control}
To maintain an appropriate and engaging level of challenge, the AI dynamically adjusts its difficulty based on player performance by varying the number of MCTS iterations and the Minimax search depth. This mechanism is not intended to optimize opponent strength, but to stabilize gameplay difficulty across participants. The AI never makes deliberate mistakes in forced win or forced block situations, preserving core tactical constraints. By reducing performance variance due to opponent mismatch, this design allows analysis to focus on participants' individual decision-making under repeated, irreversible commitments.

\subsection{Move Quality Scoring}
Within this controlled interaction environment, the system evaluates the quality of each individual player's decision. For every player move, the system computes a continuous move quality score between 0 and 1, reflecting the quality of the submitted move relative to all legal alternatives available in the same game state.

Move quality is computed using Minimax search (depth 4). For each legal move, we compute its Minimax evaluation, assign the optimal move a score of 1, and score other moves by averaging (1) a rank-normalized component based on the move's rank among legal alternatives and (2) a value-normalized component computed as $(v - v_{\min})/(v_{\max} - v_{\min})$, where $v$ is the move's evaluation and $v_{\max}$/$v_{\min}$ are the best/worst evaluations among legal moves in the same state.

\section{User Study Methods}
\label{sec:methods}

\subsection{Study Design, Participants, and Randomization}

Our IRB-approved study used a within-subject design implemented in a blockchain-based Connect Four game.
Participants were recruited worldwide from the X community via convenience sampling and required prior experience with blockchain wallets.
To be included in the analysis, participants had to provide informed consent, watch a short Connect Four tutorial video explaining the game's rules, complete at least 4 games, and complete a post-test survey.
Participants were compensated based on time spent playing and the number of wins.

\begin{figure*}[t!]
    \centering
    \includegraphics[width=1\linewidth]{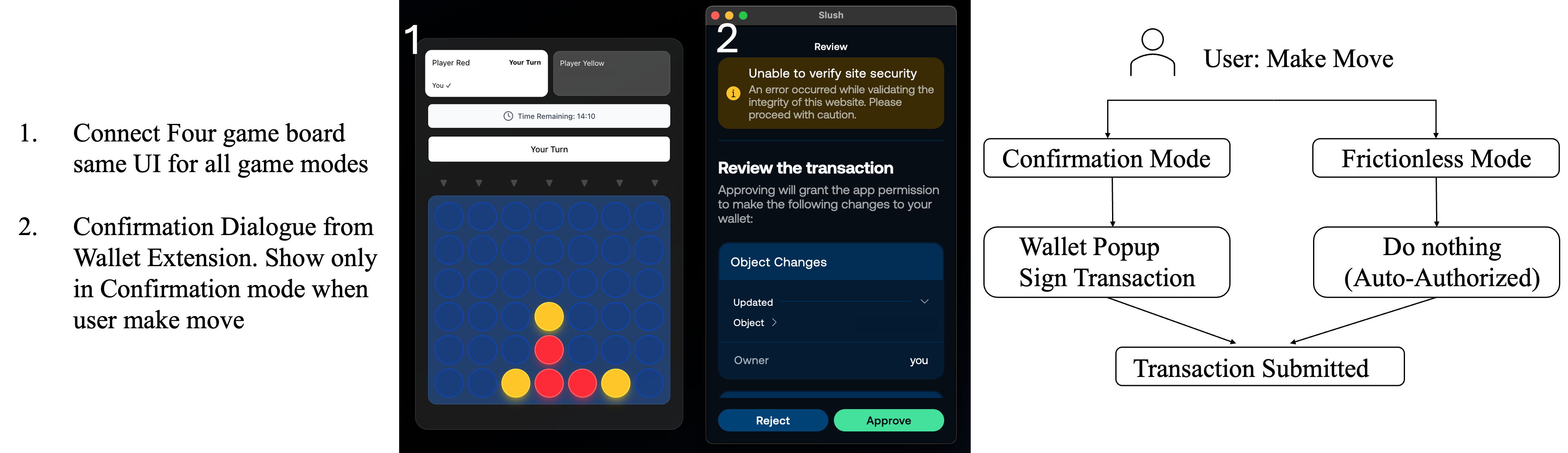}
    \caption{The figure illustrates the gameplay interface and representative screenshots, along with the user flow experienced by participants. The only difference between confirmation mode and frictionless mode is the former's wallet signature step.}
    \label{fig:screenshot_and_data}
\end{figure*}

Participants experienced two interaction modes that differed only in transaction authorization flow while keeping the game UI and transaction content identical, as shown in \autoref{fig:screenshot_and_data}.
In the \textbf{Confirmation Mode}, each move required manual wallet confirmation via a browser extension pop-up.
In the \textbf{Frictionless Mode}, moves were automatically authorized using an in-browser private key or session capability authorized at the beginning of gameplay.

Data were collected in two waves.
\textbf{Wave 1} used per-game randomization (50\% probability per mode) without enforcing counterbalancing, resulting in unequal exposure in two modes.
\textbf{Wave 2} enforced a strictly counterbalanced schedule across the first four games, ensuring complete within-subject exposure.
Across both waves, 126 participants were enrolled (Wave 1: 82; Wave 2: 44).
Of these, 109 participants met the inclusion criteria (Wave 1: 70; Wave 2: 39).

\begin{table*}[t]
\centering
\small
\begin{tabular}{p{0.48\linewidth}cccc}
\hline
\textbf{Survey Item (7-point Likert)} & \textbf{Confirmation Mode} & \textbf{Frictionless Mode} & \textbf{$p$-value} & \textbf{Cohen's $d$} \\
\hline
The interaction flow made me more likely to make mistakes while playing this version.
& 4.77 & 3.63 & 0.0007 & 0.34 \\

I feel like I have a better performance while playing this version.
& 3.61 & 5.48 & $<.0001$ & -0.61 \\

I frequently had to switch my attention between planning my next move and handling confirmations while playing this version.
& 5.34 & 3.27 & $<.0001$ & 0.65 \\

I felt frustrated or annoyed while playing this version.
& 4.54 & 2.57 & $<.0001$ & 0.62 \\

% I had to invest a lot of mental effort to stay focused while playing this version.
% & 4.96 & 4.54 & 0.1917 & 0.13 \\
\hline
\end{tabular}
\caption{Paired $t$-test results comparing subjective experience between Frictionless Mode and Confirmation Mode using full survey item text aggregated across two study waves ($N=109$ eligible respondents) in 7-point Likert-scale items. Higher values indicate stronger agreement.}
\label{tab:survey-preference}
\end{table*}

\subsection{Measures}
% We focus on three primary measures.

\textbf{Subjective experience.}  
After four rounds of gameplay, participants completed a post-test survey comparing the two modes on perceived usability, attention, error likelihood, and performance, using paired 7-point Likert-scale items. The items are listed in \autoref{tab:survey-preference}. Survey responses reflect participants' subjective perceptions of the interaction modes and do not require equal exposure to both. Survey responses were analyzed at the participant level using paired $t$-tests.

\noindent\textbf{Objective performance.}
Performance was measured using win rate and move quality. Win rate captures game-level outcomes, while move quality is a continuous score computed by the AI engine that reflects how optimal a submitted move was relative to all legal alternatives in the same state. To reduce heterogeneity from unequal play lengths and learning effects, analyses focus on each participant's first four games before the post-test survey. For inferential testing, we computed each participant's mean win rate and mean move quality separately for each mode and compared modes using paired $t$-tests, including only the data from the first four games for eligible participants.

\noindent\textbf{Pre-submission self-correction.}  
In the Confirmation mode, participants could cancel a move after opening the transaction confirmation dialog by rejecting the transaction signing request and selecting a different move in the same turn.
We refer to such cases as rejected moves. Each rejected-move episode consists of two actions taken in the same game state: the canceled action and the subsequently submitted correction. For each rejected move, we compared the move quality of these two actions using paired $t$-tests to analyze pre-submission reconsideration behavior.

Given the stronger internal validity of the counterbalanced design, our primary behavioral comparison (win rate and move quality) focuses on \textbf{Wave 2}.
We use \textbf{Wave 1} to verify consistency in effect direction.
Subjective experience (post-test survey) and independent events (rejected-move self-correction) results are aggregated across \textbf{both waves} to increase sample size and statistical power.

\section{User Study Results}
\label{sec:results}

\subsection{Participants Prefer Frictionless mode and Perceive Better Performance}
Table~\ref{tab:survey-preference} summarizes paired comparisons of subjective experience between \textbf{Frictionless and Confirmation modes}. Participants consistently reported better perceived performance, fewer interaction-induced mistakes, reduced attention switching, and lower frustration when using \textbf{Frictionless mode} than when using \textbf{Confirmation mode} (all $p < .001$).
% Notably, perceived mental effort to stay focused did not differ significantly between modes, suggesting that the subjective benefits of \textbf{Frictionless mode} are not driven by overall cognitive load reduction, but by changes in interaction flow and attentional demands.
Overall preference ratings strongly favored \textbf{Frictionless mode}, indicating that users found the automated interaction flow smoother and more effective, despite later evidence of degraded objective performance.

\subsection{Frictionless Mode Is Associated with Worse Objective Performance}
\label{subsec:game-result-performance}

\begin{table*}[t]
\centering
\small
\begin{tabular}{lccccc}
\hline
\textbf{Performance Measure} &
\textbf{Frictionless Mode} &
\textbf{Confirmation Mode} &
\textbf{Difference} &
\textbf{$p$} &
\textbf{Cohen's $d$} \\
\hline
Win rate (first four games)
& 9.5\%
& 21.2\%
& $-11.8$\%
& 0.044
& $-0.33$ \\

Move quality score (first four games)
& 0.647
& 0.698
& $-0.051$
& 0.022
& $-0.38$ \\
\hline
\end{tabular}
\caption{Objective performance comparison between Frictionless Mode and Confirmation Mode for Wave 2 ($N$ = 39).}
\label{tab:game-result-wave2}
\end{table*}

\begin{table*}[t]
\centering
\small
\begin{tabular}{lcccccc}
\hline
\textbf{Game Period} &
\makecell{\textbf{Original}\\\textbf{Rejected Move}} &
\makecell{\textbf{Submitted}\\\textbf{Correction Move}} &
\textbf{Improvement} &
% \textbf{Improved (\%)} &
\textbf{$p$ (2-tailed)} &
\textbf{Cohen's $d$} \\
\hline
First four games (Wave 1 + Wave 2)
& 0.616
& 0.737
& +0.121
% & 45.5\%
& 0.0052
& 0.36 \\
\hline
\end{tabular}
\caption{Two-tailed paired $t$-test comparing move quality scores between original rejected moves and submitted correction moves in the \textbf{Confirmation Mode}, aggregated from participants' first four games ($N=66$ paired comparisons from 35 unique participants).}
\label{tab:rejected-first4-agg}
\end{table*}

Using the counterbalanced Wave 2 dataset (first four games), we find that removing confirmation is associated with worse objective performance.
Participants achieved a lower win rate in \textbf{Frictionless Mode} (9.5\%) than in \textbf{Confirmation Mode} (21.2\%), a difference of $-11.8$ percentage points ($p=0.044$, $d=-0.33$).
Decision quality as measured by AI-evaluated move quality was also lower in \textbf{Frictionless Mode} ($M=0.647$) than in \textbf{Confirmation Mode} ($M=0.698$), with a mean difference of $-0.051$ ($p=0.022$, $d=-0.38$).
Together, these results indicate a reliable reduction in objective performance when the confirmation checkpoint is removed.
Although Wave~1 did not enforce strict counterbalancing, participants who experienced both modes ($N=45$) showed consistent effect directions with Wave~2, with lower win rate ($-7.9$\%, $p=0.085$, $d=-0.21$) and move quality ($-0.048$, $p=0.019$, $d=-0.29$) in \textbf{Frictionless Mode}.

\subsection{Rejected Moves Enable Effective Self-Correction in Confirmation Mode}
\label{subsec:rejected-moves}

To investigate a potential mechanism underlying the observed performance gap, we analyzed rejected moves in the \textbf{Confirmation Mode}.
A rejected move occurs when a participant initiates a transaction by selecting a column and opening the confirmation dialog, but then cancels the submission and selects a different column for the same turn.
We refer to the initially selected but canceled action as the \emph{original rejected move}, and the subsequently chosen and submitted action as the \emph{submitted correction move}.
These two actions represent alternative decisions made within the same game state.
Aggregating rejected-move episodes across both deployment waves and focusing on participants' first four games, we identified $N=66$ paired comparisons from 35 unique participants (32.1\% of all eligible participants, $35/109$).
As shown in \autoref{tab:rejected-first4-agg}, mean move quality increased from 0.616 for original rejected moves to 0.737 for submitted correction moves, corresponding to an average improvement of $+0.121$.
This improvement was statistically significant based on a two-tailed paired $t$-test ($p=0.0052$, Cohen's $d=0.36$), indicating that when participants canceled a submission and reconsidered, they tended to select a higher-quality move.
In 45.5\% of cases (30/66), the correction move had a higher quality score than the rejected move; in 31.8\% (21/66), the two moves had equal quality scores; and in 22.7\% (15/66), the correction move had a lower quality score.

\section{Discussion and Conclusion}

Our results reveal a dissociation between subjective experience and objective performance in blockchain authorization workflows. Participants preferred frictionless, automated authorization and believed they performed better, yet objective decision quality declined when transaction confirmation was removed. This dissociation highlights a fundamental tension between usability optimization and responsible decision-making in irreversible interaction contexts.

\subsection{Authorization automation as a decision-quality trade-off}

A common assumption in user experience design is that reducing friction is beneficial \cite{voskobojnikov2021u,albayati2021study}. Our findings challenge this view by demonstrating a difference between perceived usability and decision quality. Although authorization automation improves interaction smoothness and users' confidence, it can also alter how users allocate attention and deliberation. In our study, manual confirmation was associated with better objective performance, despite being perceived as more interruptive and less usable.

This dissociation is consistent with prior work on automation-related overreliance and complacency \cite{parasuraman1997humans,parasuraman2010complacency,dzindolet2003role}. By removing explicit checkpoints, automated systems may encourage faster commitments and reduced monitoring. Importantly, our results do not suggest that automation is inherently harmful. Rather, they identify a boundary condition that is particularly salient for authorization workflows involving irreversible actions, where removing the commitment boundary may improve perceived efficiency while undermining decision quality.

\noindent\textbf{Implication on responsible friction and calibrated confidence:}
Our results reveal a responsibility gap in systems that optimize solely for perceived usability. Users prefer frictionless interactions and report higher confidence, yet their objective decision quality may decline when explicit confirmation is removed. We therefore argue for \emph{responsible friction}: interaction designs that preserve smooth interaction flow while reintroducing lightweight opportunities for verification at commitment boundaries. A central challenge is \emph{calibrated confidence} that better aligns users' perceived performance with their actual decision quality, potentially through context-sensitive or risk-aware authorization mechanisms.

\subsection{Beyond binary: toward a spectrum of confirmation designs}

Our experimental design deliberately compared two extremes, full manual confirmation versus fully automated authorization, to isolate whether the presence of a commitment boundary affects decision quality at all. We acknowledge that this binary framing does not capture the full range of confirmation designs used in practice. Real-world systems employ a variety of intermediate approaches, including brief transaction previews, non-blocking visual warnings, risk-highlighted summaries, and progressive disclosure of transaction details, that aim to balance usability with user awareness.

We view our study as establishing a necessary first step: demonstrating that the \emph{presence} of a deliberation boundary matters for decision quality. Having established this baseline, the natural next question is how much friction is needed and in what form. Future work should systematically vary the salience and intrusiveness of confirmation cues to identify where on this spectrum the benefits of reflection can be preserved while minimizing usability costs. For instance, context-sensitive designs might present lightweight confirmations for routine, low-risk transactions while reserving full confirmation dialogs for high-value or unusual actions.

\subsection{Confirmation as a reflective checkpoint rather than a disruptive interruption}

At first glance, these findings may appear to conflict with prior interruption research showing that interruptions degrade performance \cite{altmann2014momentary,monk2004recovering,hodgetts2006interruption}. However, unlike externally imposed interruptions, transaction confirmation in our system is directly triggered by the user's own action and occurs immediately before an irreversible commitment.

One interpretation is that confirmation functions as a reflective checkpoint at a natural commitment boundary, prompting reconsideration or heightened awareness of irreversibility. The underlying mechanism likely involves multiple factors: the additional time created by the confirmation step, the psychological salience of an explicit signing action that signals irreversible commitment, and an attentional reset that disrupts autopilot-like interaction patterns. Our rejected-move analysis provides direct behavioral evidence for one such mechanism: when participants encountered the confirmation dialog, they sometimes recognized a suboptimal choice and self-corrected before commitment. Although only 33\% of participants exhibited explicitly rejected moves, overall performance remained higher in the \textbf{Confirmation Mode}, suggesting that confirmation may support reflection through subtler mechanisms beyond observable cancellation behavior, such as increased attentiveness or more deliberate evaluation of the chosen action.

\subsection{Ecological validity and transferability}
\label{subsec:ecological}

A key question is whether findings from a strategy game generalize to real-world blockchain transactions. We acknowledge that the cognitive demands differ: Connect Four moves involve strategic calculation errors (failing to foresee opponent traps), whereas real crypto transactions typically involve verification errors (overlooking gas fees, misreading permission scopes, or falling for phishing attacks). The nature of ``mistakes'' in these two domains is fundamentally different.

However, we argue that the core mechanism our study identifies, that an explicit commitment boundary provides an opportunity for deliberation and self-correction before an irreversible action, is domain-general. Whether a user is reconsidering a game move or double-checking a token transfer amount, the confirmation dialog creates a pause that can interrupt hasty commitment. Our game-based design maximizes internal validity by isolating this single variable (the presence of a confirmation step) while holding all other factors constant, something that would be difficult to achieve in a field study with real financial stakes, heterogeneous transaction types, and varying risk levels.

Validating these findings in higher-stakes environments remains an important direction. Future studies could examine confirmation effects in decentralized finance (DeFi) trading interfaces, governance voting, or NFT purchases, where the stakes are real and the informational complexity is greater. Such studies would also need to account for the Web3-specific challenge that users often struggle with technical terminology (e.g., gas fees, smart contract approvals), which may amplify the value of confirmation as an opportunity to parse unfamiliar information. We emphasize that our study establishes the \emph{cognitive role} of commitment boundaries, rather than evaluating specific wallet UI designs.

\subsection{Limitations and future work}

Beyond ecological validity (discussed above), several additional limitations should be noted. First, our participant sample was recruited from X (formerly Twitter) blockchain communities and required prior wallet experience, which may limit generalizability to novice users or broader populations. An open question is whether lower-skill or less experienced participants benefit more from confirmation steps than expert users, which future work could investigate through skill-stratified analyses. Second, we did not measure process-level mechanisms such as per-move thinking time, which could help disentangle whether the performance benefit arises from additional deliberation time, attentional reframing, or the psychological weight of the signing action. Third, this study compared two extreme points on the friction spectrum. As discussed, future work should systematically explore intermediate designs, such as adaptive confirmation that adjusts intrusiveness based on transaction risk, low-interruption micro-checkpoints (e.g., brief visual highlights before commitment), and context-aware delegation boundaries that preserve smooth flow for routine actions while flagging unusual ones.

\subsection{Conclusion}

Authorization automation plays a central role in making blockchain applications more accessible and usable. Our results show that removing transaction confirmation improves subjective user experience but can reduce objective decision quality. Rather than viewing friction solely as a usability cost, our findings suggest that confirmation steps can function as cognitively meaningful checkpoints. Designing usable and responsible blockchain systems may therefore require balancing interaction smoothness with opportunities for reflection, particularly in irreversible decision contexts.

% \begin{acks}
% TBA
% \end{acks}
\bibliographystyle{ACM-Reference-Format}
\bibliography{reference}

% \appendix
% \section{Appendix: Screenshot of the system}

% \begin{figure}[h]
%     \centering
%     \includegraphics[width=1\linewidth]{images/image.png}
%     \caption{The screenshot of the system}
%     \label{fig:placeholder}
% \end{figure}

% \appendix

\end{document}